# FAULT-TOLERANT MULTIPATH ROUTING SCHEME FOR ENERGY EFFICIENT WIRELESS SENSOR NETWORKS

PrasenjitChanak,TuhinaSamanta,Indrajit Banerjee

Department of Information Technology
Bengal Engineering and Science University, Shibpur, Howrah-711103, India
prasenjit.chanak@gmail.com, t_samanta,ibanerjee,@it.becs.ac.in

***ABSTRACT***—*Themain challengein wireless sensor network is to improve the fault tolerance of each node and also provide an energy efficient fast data routing service. In this paper we propose an energyefficient node fault diagnosis and recovery for wireless sensor networks referred as fault tolerant multipath routing scheme for energy efficientwireless sensor network (FTMRS).The FTMRSis based on multipath data routing scheme. One shortest path is use for main data routing in FTMRS technique and other two backup paths are used as alternative path for faulty network and to handle the overloaded traffic on main channel.Shortest path data routing ensures energy efficient data routing. The performance analysis of FTMRSshows better results compared to other popular fault tolerant techniques in wireless sensor networks.*

## 1. INTRODUCTION

Wireless sensor network is a collection of hundreds and thousands of low cost, low power smart sensing devices. Sensor nodes are deployed in a monitoring area. They collect data from monitoring environment and transmit to base station (BS) by multi-hope or single hope communication. In WSN, fault occurrence probability is very high compare to traditional networking [1]. On the other handnetworks maintenance and nodes replacement is impossible due to remote deployment. These features motivate researchers to make automatic fault management techniques in wireless sensor networks. As a result now a day's different types fault detection and fault tolerance techniquesare proposed [2], [3]. Kim M, et al., proposed a multipath fault tolerant routing protocol based on the load balancing in 2008 [4]. In this paper, authors diagnose node failures along any individual path and increase the network persistence. The protocol constructs path between different nodes. Therefore, protocol leads to high resilience and fault tolerance and it also control message overhead. Li. S and Wu. Z proposed a node-disjoint parallel multipath routing algorithm in 2006[5]. This technique uses source delay and onehop response mechanism to construct multiple paths concurrently. In 2010 Yang Y. et al., [6] proposed a network coding base reliable disjoint and braided multipath routing. In this technique the authors construct disjoint and braided multipath to increase the network reliability. It also uses network coding mechanism to reduce packet redundancy when using multipath delivery. Y. Challal et al., proposed secure multipath fault tolerance technique known as SMRP/SEIF in 2011[7]. This technique introduces fault tolerant routing scheme with a high level of reliability through a secure multipath communication topology. Occurrences of fault in wireless sensor network are largely classified in two groups; (i) transmission fault,and (ii) node fault. The node fault [8], [9], [10] is further classified into five groups. These arepower fault, sensor circuit fault, microcontroller fault, transmitter circuit fault and receive circuit fault as discussed in [11].Energy efficiency is a prime metric in WSN performance analysis. This motivates us to propose an algorithm for fault tolerant energy efficient routing.

In this paper, we propose a fault tolerant routing which involves fault recovery process with fault detection scheme, referred to as energy efficient fault tolerantmultipath routing scheme for wireless sensor network (FTMRS). In FTMRS technique every sensor node transmitsits data to a base station through shortest path. If data or node fault occurs in the network, these are recovered very fast.The data are transmitted to base station withminimum time and energy loss. The FTMRS also controls the data traffic when data are transmitted to cluster head or base station(BS).

The rest of this paper is organized as follows. Section 2 describe proposed load balanced model.In section 3, we propose architecture for FTMRS. The proposed methodology for FTMRS is discussed in section 4. Performances and comparison result are showed in section 5. Finally, the paper is concluded in section 6.

## 2. PROPOSED LOAD BALANCING MODEL

In FTMRS technique, we use standard data communication model originally proposed in [12]. In FTMRS technique cluster size are calculated with the help of*theorem 1* and *theorem 2*. *Theorems 1*establishes a relation between number of message passing through a node and nodes energy.The *theorems 2* establish a relation between numbers of nodes connection of a particular node with number of message passing in a particular time. The load of a node is directly affected by the number of node connected to it. If number of node connection is increased then load on that particular node isincreased. On the other hand if load of a node is increased, energy loss of the sensor node is increased.

**Definition 1:** The load$P_j$of a node is depending on number of data packet receives and transmitsby a particular node. The data load on a particular node is depends on number of sensor nodes connected with it andamount of data sensed by this particular node. The $S_p$ denote a data packet receive by a single connection and$S_d$denote a data packet transmitby single connection.$P_j = \sum_{i=0}^{n} S_p + S_d$

**Theorem 1:**If initial energy of a sensor node is U, then partial derivative of the total energy of a sensor node expressed in terms of number of message passing $\lambda_j$ at node *j* is equal to the load$P_j$ at sensor node *j*. This theorem expressed symbolically as $\frac{\partial U}{\partial \lambda_j} = P_j$

**Proof: -**Consider a series of loads $P_1, P_2, P_3, \dots P_j, \dots P_n$are acting on node 1, 2... j, ..., n who areproducing number of message $\lambda_1, \lambda_2 \dots \lambda_j, \dots \lambda_n$.

Now impose a small increment $\delta\lambda_j$ to the message passing at the node *j* keeping all other load unchanged. As a consequence, the increments in the loads are$\delta P_1, \delta P_2, \dots \delta P_j, \dots, \delta P_n$. The increment in the number of message at node*j* and consequent increments in loadsat all the neighbour nodes. Therefore,$P_j \delta\lambda_j = \lambda_1 \delta P_1 + \lambda_2 \delta P_2 + \cdots . + \lambda_j \delta P_j + \cdots + \lambda_n \delta P_n$,$\frac{\delta U}{\delta \lambda_j} = P_j$In the limit $\delta\lambda_j \to 0$, the above equation becomes $\frac{\partial U}{\partial \lambda_j} = P_j$

**Theorem 2:** Partial derivative of the energy loss in sensor nodes expressed in terms of load with respect toany load$P_j$ at any sensor nodes j is equal to the number of message passing $\lambda_j$through the*j^th^* node. This theorem may be expressed mathematically as $\frac{\partial U}{\partial P_j} = \lambda_j$

**Proof: -**Consider a series of loads $P_1, P_2, P_3, \dots P_j, \dots P_n$ acting on a node j and for this section the messagepassed are $\lambda_1, \lambda_2 \dots \lambda_j, \dots \lambda_n$. Now impose a small increment $\delta P_j$ to the load at the node j keeping all other factors unchanged.As a consequence, the message passing increasesby$\delta\lambda_1, \delta\lambda_2, \dots \delta\lambda_j, \dots, \delta\lambda_n$. However, due to increments in the load at node *j*,there is a consequent increment in message passing in all the neighbouring nodes. Therefore,$P_1 \delta_1 +$ $P_2\delta_2 + \cdots + P_j\delta_j + \cdots + P_n\delta_n = \lambda_j\delta P_j$ , $\frac{\delta U}{\delta P_j} = \lambda_j$ in the limit $\delta P_j \to 0$, the above equation becomes $\frac{\partial U}{\partial P_j} = \lambda_j$.

## 3. ARCHITECTURE FOR ENERGY EFFICIENTFAULT TOLERANT MULTIPATH ROUTINGSCHEME (FTMRS):

In FTMRS technique, cluster sizes are calculated based on the cluster head load using theorem 1 and theorem 2. The clusters head load depends on the number of message received in cluster head and number of data transmitted from cluster head.

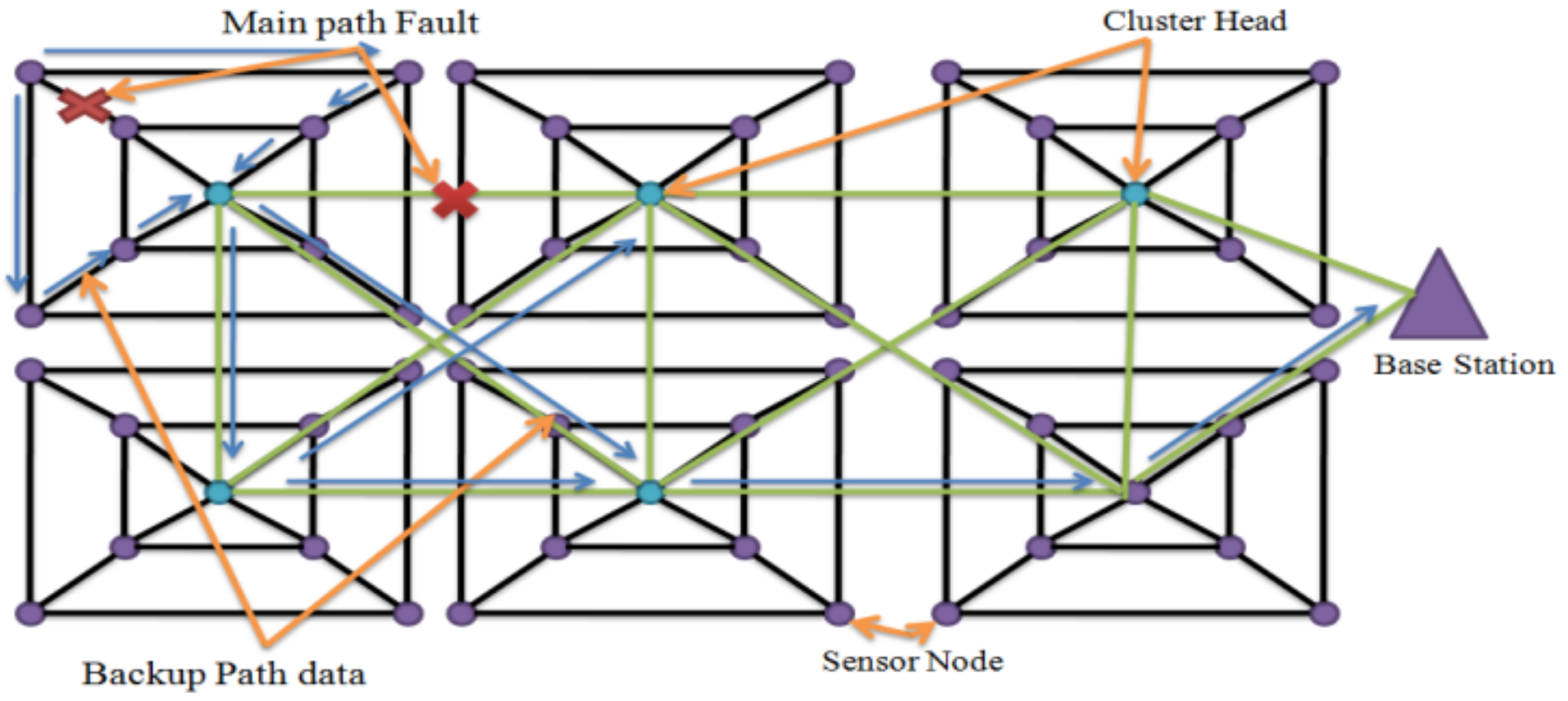

**Figure1:** Alternative path data routing in FTMRS

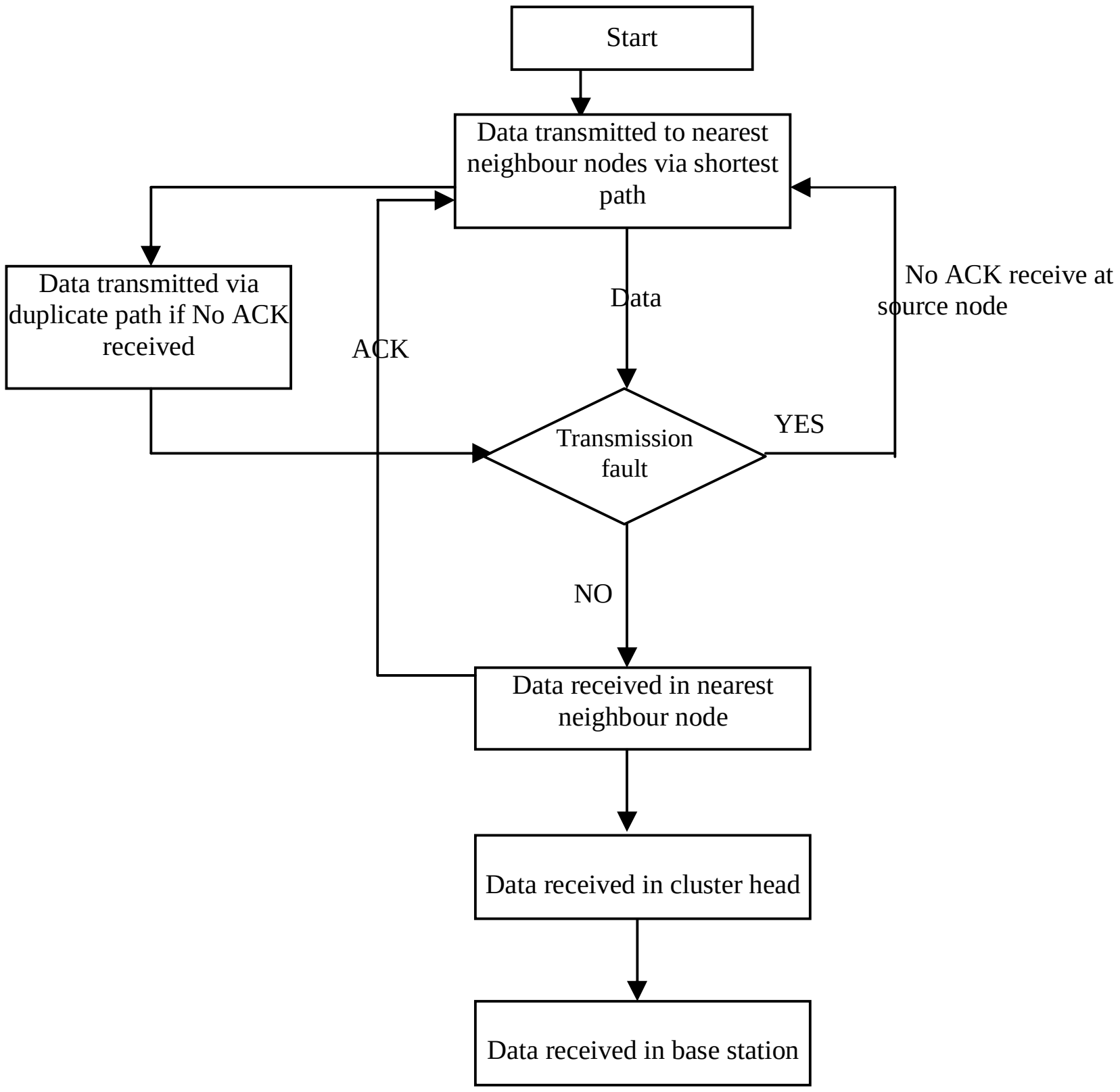

Figure 2: FTMRS architecture for fault tolerance

### 3.1 Fault Tolerance Data Routing Model

In FTMRS technique sensor nodes are arranged into small clusters. Every cluster contains a cluster head and cluster member node. Every cluster member node is capable of sending data over multiple paths. Cluster member nodes send their data in a shortest path to CH.Other paths are used for duplicated data transmission. Within a cluster a cluster member node sends its data to other cluster member nodesthrough three alternative paths. One of them is shortest, which is responsible for fast data transmission to cluster head. However, due to any external or internal problem shortest path fails,then next available shortest alternativepathis used to recover the faulty data transmission (Figure1).

In FTMRS technique when data are reaching to the neighbouring destination node via data routing path, they first check their received data and their own sensed data. If these two are same then neighbour nodes are not forwarding the received data to others. If a node receives different data then receiver node sends receiving data toward the cluster head with shortest path.

In FTMRS technique, clusters head and base station arealso connected to each other with the help of multiple (three) data path(Figure1). The shortest path is mainly used for fast energy efficient data routing towards base station. Other two backup paths are used for duplicate data routing, which makes the network path fault tolerant.

## 4. PROPOSED METHOD FOR FTMRS

In this section, we briefly describe our proposedFTMRS technique. This section is divided into two sub section one is fault tolerant data routing another is Energy efficient routing methodology.

### 4.1 Fault Tolerance Data Routing

In FTMRS technique, every cluster member node transmits data to cluster head.Cluster head collects all cluster member data. Cluster head after data aggregation transmits to base station.

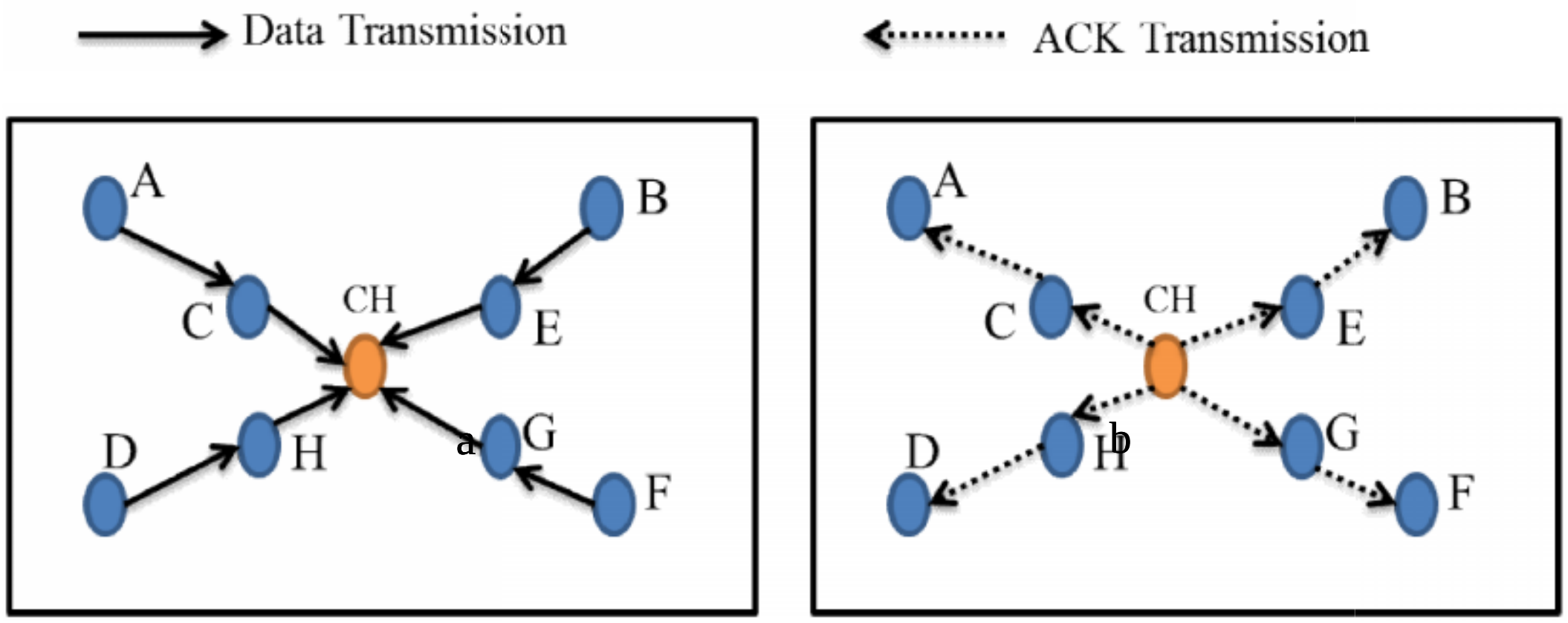

**Figure3:** Data transmission policy in FTMRS technique

In Figure 3.a. Cluster member nodes A, B, D, F transmits their data to cluster head’s nearest to cluster member nodes C, E, H, and G respectively. If any node failure or transmission fault does not occurs, then after data is receivedby the cluster head’s nearest member nodes C, E, H, and G send acknowledgement messages (Figure 3.b).In the same way, cluster head nearest member nodes C, E, H and G send their data to cluster head CH.

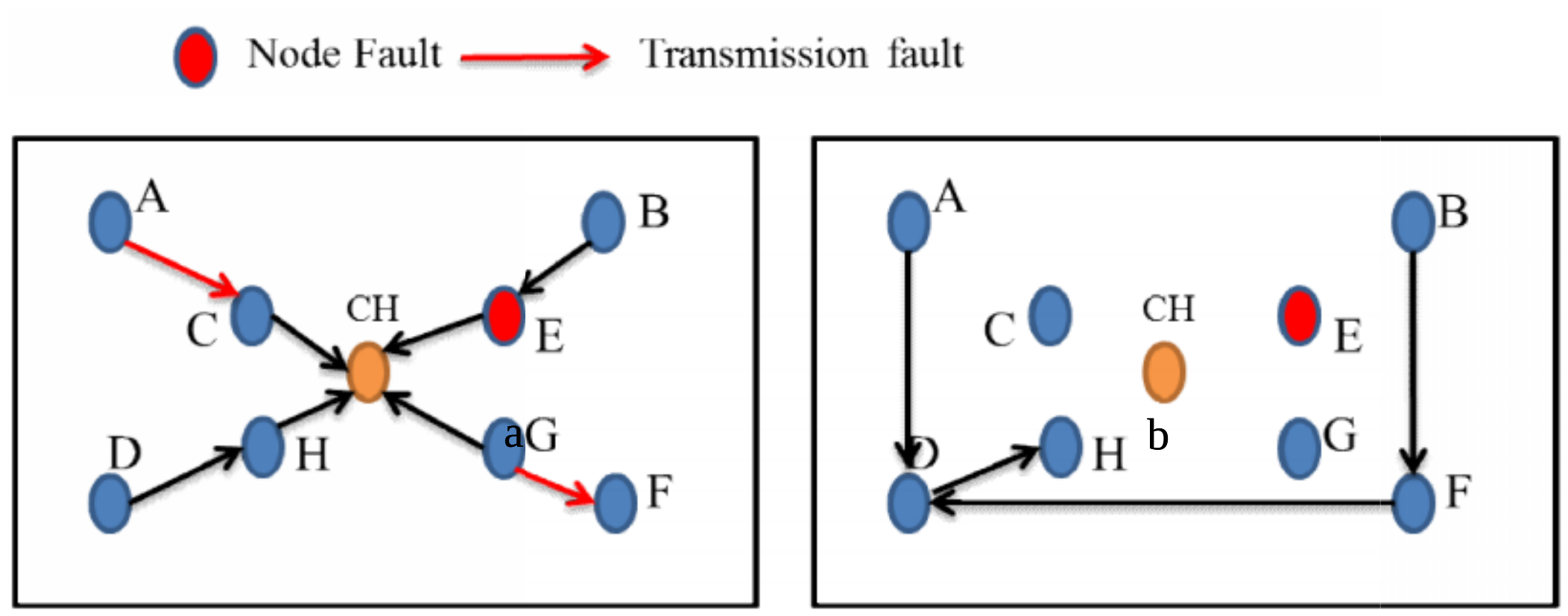

**Figure4:** Fault recovery policy in FTMRS technique

In the FTMRS technique if any transmission fault or node fault occurs in the network, then source nodes A, B, D or F does not receive any acknowledge message. Therefore, they are sending their data through duplicate path. In Figure4.a A to C node transmission fault occurs, hence'A' retransmitsits sensing data to D node. Similarly, when node E fails, node 'B'transmitsits data to 'F' node. However, node F does not gate any acknowledge message from G; hence it retransmits its data to node D (Figure4.b). In the same way cluster head transmits their data to the base station in FTMRS technique.

### 4.2. Node Hardware Fault Detection

In FTMRS scheme,the node collects data from nearest neighbour and transmits to the cluster head by a shortest path. If a node is not receiving any data from its neighbouring node for a period of time then the node sendsa health message to neighbour node and waits for replay message. If all neighbour nodes replay with respect to that health message, then node decides a transitions fault occurs in previous transmission. On the other hand if a node does not receive any replay message against health message then the node decidesits receiver circuit is faulty. However, if any one of the neighbour node is not replaying against health message then the node decidesthat the transmitter circuit of that neighbouring node is faulty. Then it informsthis to all other neighbour nodes. Sensor circuit fault is detected by the node itself by comparing its sensed data with data that he has been received from neighbour node. The comparison technique used here is explained in detail in [11].If sensing information is less thanthe threshold value, then the node's sensor circuit is in active condition. If sensing information is grater then the threshold value, then the sensor circuit is faulty. The FTMRS fault detection algorithm is described below.

**Algorithm 1: Fault Detection Algorithm**

**Input:** Insert all nodes into S (array of nodes)
**Output:** Check nodes hardware condition and find out fault nodes

**WHILE** S! =Null **DO**
**WHILE** network is alive **DO**
**FOR** each node **DO**
    **IF** node receive data from neighbour **THEN**
        Receiving data transmitted to shortest path
    **ELSE**
        Send health message all neighbour nodes
    **IF** receive replay all neighbour nodes **THEN**

```
                Transmission fault occurs in previous transmission
        ELSE IF not receives replay from communication node THEN
                Communication node is dead.
                Inform to all neighbour node.
        ELSEIF not receive replay from all neighbour THEN
                nodes receiver circuit fault
        END IF
        END IF
        IF Neighbour node data <= threshold value THEN
                Sensor circuit node is good
        ELSE
                Sensor circuit of comparison nodes is Faulty
                Inform to cluster head
        END IF
        IF node battery reading < threshold value THEN
                Battery fault occur
        ELSE
                Battery is good
        END IF
  END FOR
  END WHILE
END WHILE
```

### 4.3 Fault Recovery in FTMRS

Depending on the hardware condition of the node they are categorize as, *Normal Node, Traffic Node, End node, Dead Node* (Table 1) [11]. The categorization helps improving the network lifetime and decreases the percentage of dead node in the network.

Table 1
Categorization of nodes with respect to different hardware circuit failure

| Node category | Microcontroller | Sensor circuit | Transmitter circuit | Receiver circuit | Battery /Power |
|---|---|---|---|---|---|
| Normal Node | Non Faulty | Non Faulty | Non Faulty | Non Faulty | Non Faulty |
| Traffic node | Non Faulty | Faulty | Non Faulty | Non Faulty | Non Faulty |
| End node | Non Faulty | Non Faulty | Non Faulty | Faulty | Non Faulty |
| Dead Node | Faulty | Faulty | Faulty | Faulty | Faulty |

The FTMRS scheme reuses the faulty sensor node depending on the node's fault condition. If the sensor circuit is faulty, then sensor node is used as a *traffic node*. On the other hand if the node's receiver circuit is faulty, then this node works as an *end node*. If a node detects its transmitter circuit, microcontroller circuit, or battery is faulty, then the node is declared as a *dead node*. In FTMRS the faulty node recovery algorithm is shown below.

**Algorithm 2: Faulty Node Recovery Algorithm**

**Input:** Sensor nodes hardware condition
**Output:** According to hardware fault condition nodes responsibility distributed

**IF** node detected sensor circuit fault **THEN**
    Declare itself as traffic node and inform all other neighbour nodes
**ELSE IF** receiver circuit fault occur **THEN**
    Declare as end node and inform all other neighbour nodes
**ELSE** transmitter circuit or microcontroller or batter fault occur **THEN**
    Declare it as dead node by the cluster head and inform other cluster member nodes
    The cluster head activate a neighbour standby node to replace the dead node.
**END IF**

### 4.4 Traffic Management Scheme

FTMRS technique deals with data traffic congestion in the network. In FTMRS, every node maintains a time interval between two different data packet transmission in the same path. A sensor node when receives a new data from other node, they first checks shortest path condition for data transmission. If the shortest path is non-faulty, and it is currently not in use, then received data is transmitted through that shortest path. However, if shortest path is in use then they transmit new data through backup path as shown in the flow diagram (Figure 2). The received data transmission technique follows the fast come fast serve (FCFS) policy.

### 4.5 Energy Efficient Routing Methodology

In FTMRS technique every node maintains a time slot for data transmission through each path.

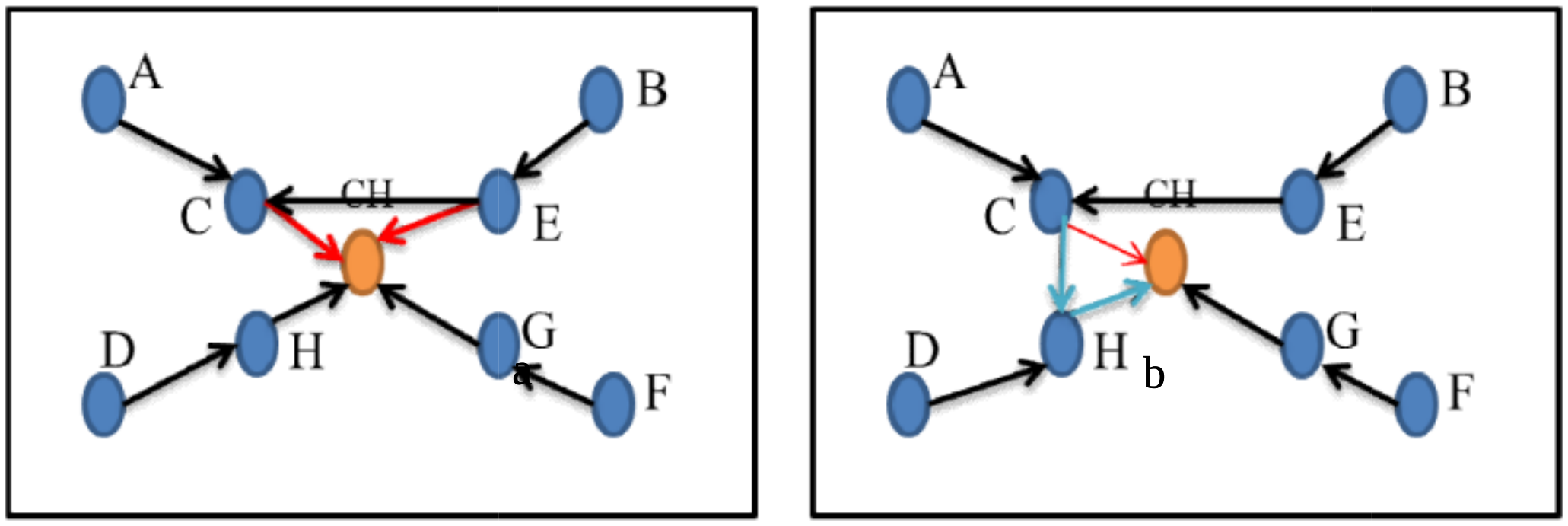

**Figure5 :**Data Traffic management in FTMRS technique

In Figure5.a when node ‘E’ transmits data to cluster head then transmission fault occurs. For this reason ‘E’ transmit data to neighbour node ‘C’. When ‘C’ receivesE’s messagethenitsends the data via available shortest path, as is shown in Figure5b, via node ‘H’. In our proposed scheme, instead of initial multipath data propagation as in [13], [14], it sends the data through a single shortest path. However, if data transmission faults occur in that path then it will send the data through alternative backup path. Therefore, the energy wastage for multipath data propagation can be saved in FTMRS.Transmission energy loss of a sensor node is $T_E$[12].

$$T_E=(\alpha_1 + \alpha_2 \square r^n) \square \beta \quad (1)$$

Where $\alpha_1$ [J/bit] is the energy loss per bit by the transmitter electronics circuit, and $\alpha_2$[J/bit/m$^4$]is the dissipated energy in the transmitter op-amp. Transmission range is$r$[m]. The parameter n is power index for the channel path loss of the antenna. $\beta$[bit]is the message size which is transmitted by each node.Receiving energy loss of a node is $R_E$[J/bit].

$$R_E=(\alpha_3) \square \beta \quad (2)$$

Where, $\alpha_3$[J/bit] is energy per bit which is consumed by the receiver's electronics circuit used by the node. $L_i$ message size which is received by each sensor node.

**Lemma 1:**The energy loss in FTMRS is less than multipath fault tolerant technique.

**Proof**:A single data communication energy loss is $E_{TR} = (T_E + R_E)$ . In multipath data transmission communication energy loss is $ME_P = \sum^n E_{TR}$. Where *n* is the number of duplicate data transmission path in multipath data routing. In FTMRS scheme *n* value is 1 on the other hand multipath fault tolerance techniques *n* is always grater then 1. Therefore, energy conservation of FTMRS is grater then to other multipath fault tolerant techniques [13], [14].

The performance analysis of FTMRS technique is discussed next.

## 5. PERFORMANCE OF FTMRS

In this section, we present the result obtained from simulating different scenarios under different network sizes, different percentage of nodes faults and transmission faults. In order to evaluate the performance of FTMRS, four traditional metrics of WSN have been considered.(i) *Global Energy of Network:*this is the sum of residual energy of each node in the network. We calculate this value at each round of data transmission. (ii) *Average delay:*Average latency from the moment of data transmitted from source node to base station. (iii) *Average packet delivery ratio*: Number of packet transmitted to the source node and number of packet receive at the destination node.(iv) *Average dissipated Energy:* Total energy loss of the network and total number of sensor nodes ratio. The simulation parameters are taken from [12], [15], [16]. The table 2 shows parameters values which are used in simulation.

Table 2: Simulation parameters

| **Parameters** | **values** |
|---|---|
| Number of node | 1000-5000 |
| Data Packet Size β | 800bit |
| Initial Energy | 0.5J |
| Energy consumed in the transmitter circuit $\square_1$ | 50 nJ/bit |
| Energy consumed in the amplifier circuit $\square_2$ | 10pJ/bit |

Table 3: EEFTMR global energy loss in 0% and 40% node failure

| Number of rounds | Global Energy (Joules/round) | | | | | | | |
|---|---|---|---|---|---|---|---|---|
| | In 0% node fault | | | | In 40% nodes fault | | | |
| | Network size (300 m× 300m) | | | | | | | |
| | Deployed Node=1000 | Deployed Node=1500 | Deployed Node=2000 | Deployed Node=2500 | Deployed Node=1000 | Deployed Node=1500 | Deployed Node=2000 | Deployed Node=2500 |
| 0 | 500 (J) | 750(J) | 1000(J) | 1250(J) | 500(J) | 750(J) | 1000(J) | 1250(J) |
| 100 | 458.35 | 696.43 | 937.50 | 1180.6 | 441.90 | 681.25 | 926.55 | 1170.60 |
| 200 | 416.7 | 642.86 | 875.00 | 1111.20 | 383.80 | 612.50 | 853.10 | 1091.20 |
| 300 | 375.05 | 589.29 | 812.50 | 1041.80 | 325.70 | 543.75 | 779.65 | 1011.80 |
| 400 | 333.40 | 535.72 | 750.00 | 972.40 | 267.60 | 475.00 | 706.20 | 932.40 |
| 500 | 291.75 | 482.15 | 687.50 | 903.00 | 209.50 | 406.25 | 632.00 | 853.00 |
| 600 | 250.10 | 428.58 | 625.00 | 833.60 | 151.40 | 337.50 | 554.30 | 774.60 |
| 700 | 204.45 | 374.01 | 562.50 | 764.20 | 93.00 | 268.75 | 485.85 | 694.20 |
| 800 | 166.8 | 321.44 | 598.00 | 694.80 | 32.20* | 200.00 | 412.40 | 614.80 |
| 900 | 125.15 | 267.87 | 437.50 | 625.40 | $0.00^{+}$ | 131.25 | 338.95 | 535.40 |
| 1000 | 83.5 | 213.30 | 375.00 | 556.00 | - | 62.50* | 265.50 | 456.23 |
| 1100 | 41.85* | 160.73 | 312.50 | 486.60 | - | $0.00^{+}$ | 192.05 | 376.60 |
| 1200 | $0.00^{+}$ | 107.16 | 250.00 | 417.20 | - | - | 118.60 | 297.20 |
| 1300 | - | 53.59* | 187.50 | 347.80 | - | - | 45.15* | 217.80 |
| 1400 | - | $0.00^{+}$ | 124.00 | 278.40 | - | - | $0.00^{+}$ | 138.40 |
| 1500 | - | - | 62.50* | 208.67 | - | - | - | 59.00* |
| 1600 | - | - | $0.00^{+}$ | 139.60 | - | - | - | $0.00^{+}$ |
| 1700 | - | - | - | 70.00* | - | - | - | - |

Table 3 depicts the global energy loss of WSN with different network size. In FTMRS, global energy lose increase when network size increase with no node failure. The life span of WSN decreases with 40% node failure, because of node fault detection and multipath data transition.

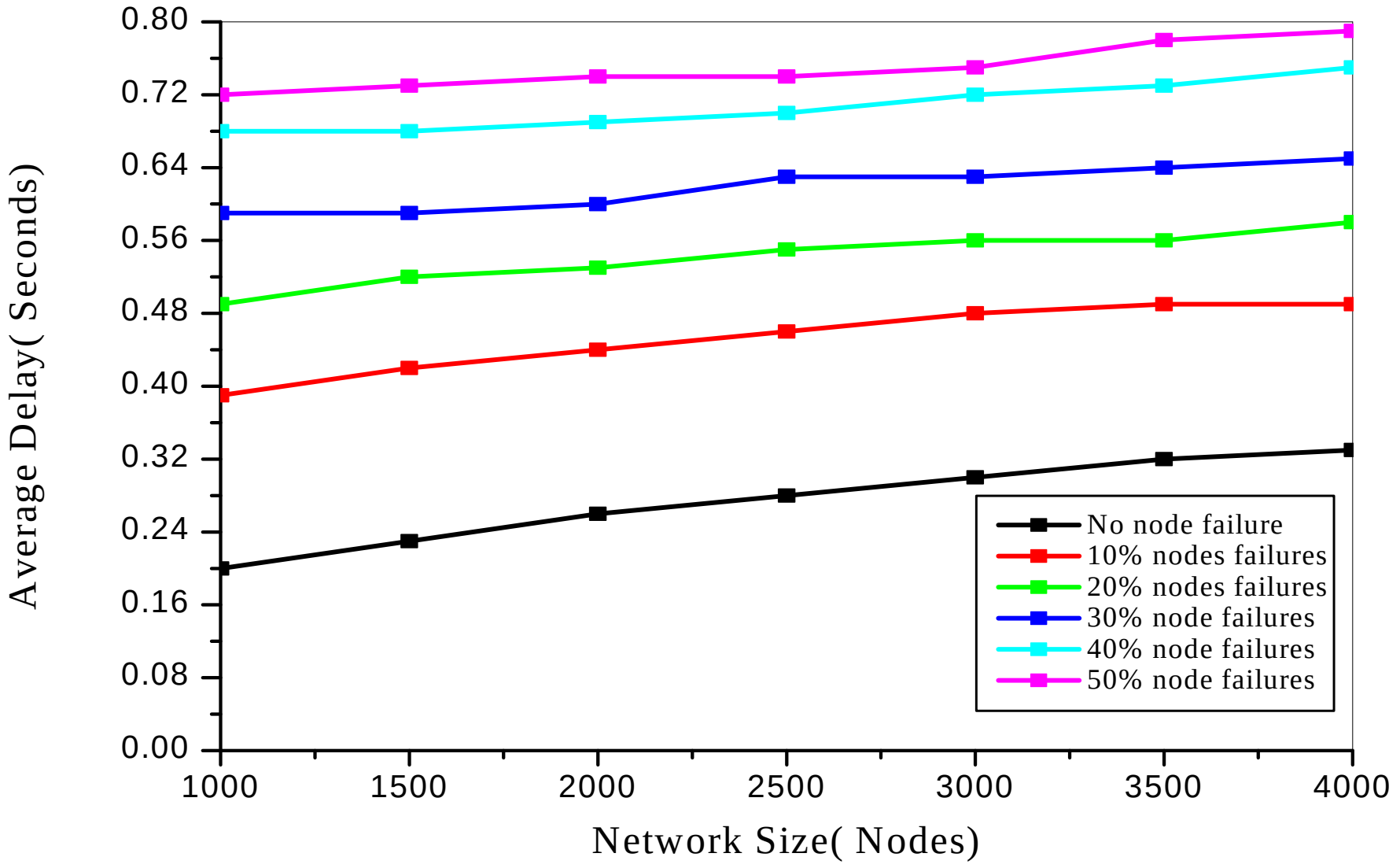

**Figure6:** Average delay in different percentage of nodes failures

Figure 6 shows the average packet transmission delay from sensor nodes to base station in different networks size. In FTMRS technique, data delivery time increase very slowly when node faults occurs. The Figure 7 shows the average packet delivery ratio from sender to base station. In the FTMRS technique, number of packet receives percentage in base station with respect to source node data transmission is very high. If any packet loss by nodes fault and path fault then backup path transmit duplicate data to cluster head as well as base station.

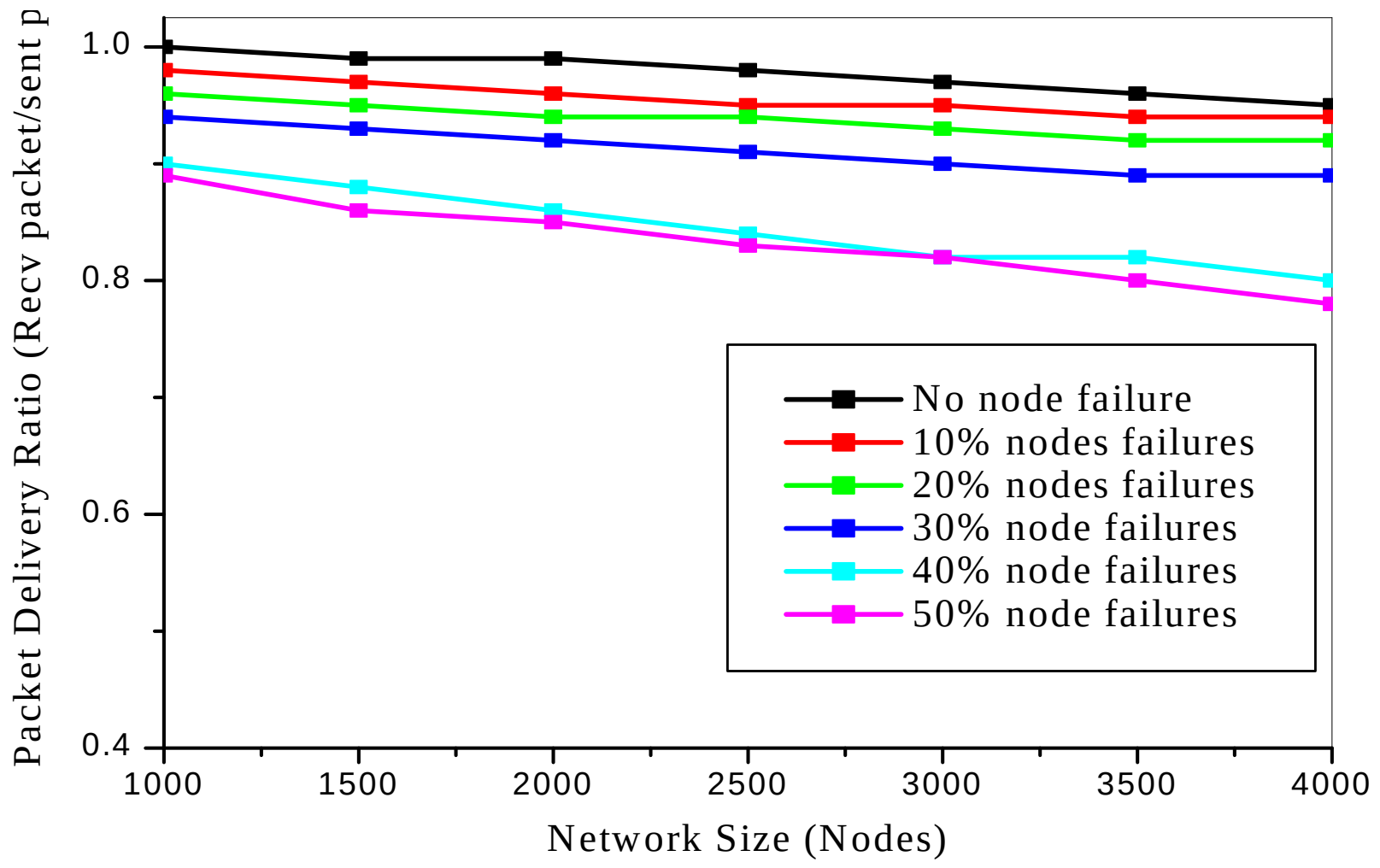

**Figure 7:** Average packet delivery in different percentage of nodes failures.

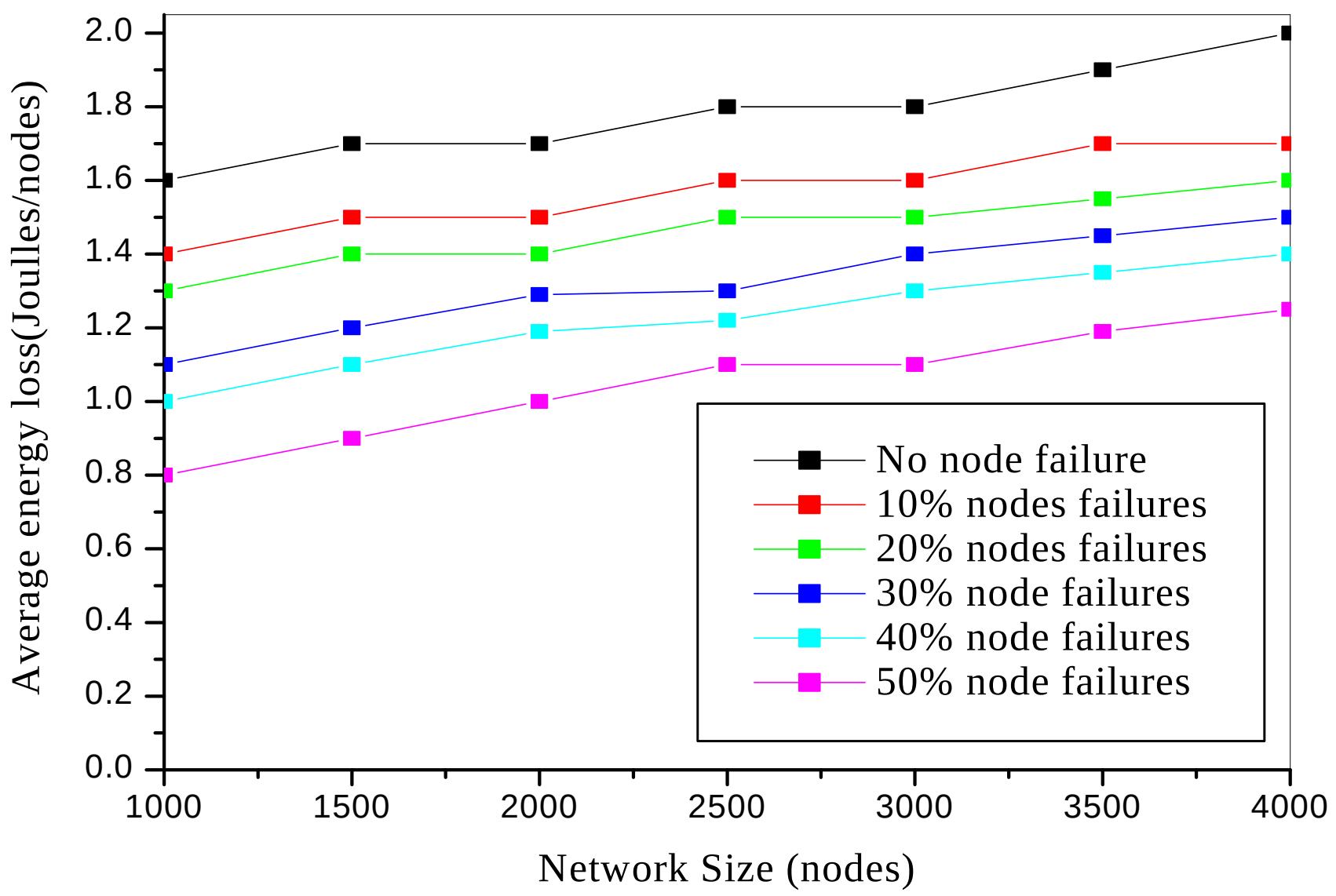

**Figure8:** Average dissipated energy in different percentage of nodes failures.

In FTMRS technique, energy loss rate of every node in different network size with different percentage of nodes failures is shown in Figure 8. When number of nodes fault percentage is low then energy loss of the networks is high because in this time maximum data is delivering to base station. If the number of node fault is increased then energy is loss of the network is decreased because in this condition data delivery to base station decreased.

Figure 9 shows the through put of the sensor nodes with respect to main routing path failures. In FTMRS technique throughput of sensor nodes is 49% batter in comparison to the fault-tolerant routing protocol for high failure rate wireless sensor networks(ENFT- AODV) [13] technique and 70% batter, compared to ad-hoc on-demand distance vector(AODV)[17]techniques.In the case of AODV technique, the throughput of the sensor nodes decreased rapidly when number of main path increases, because in this technique one main path have been failed then no other path is exited for retransmission of faulty data . ENFT-AODV technique used one backup path to improve retransmission of packet. If backup path fails then there is no way to transmit data to destination node. On the other hand FTMRS technique uses two backup paths for improvement of failure recovery.

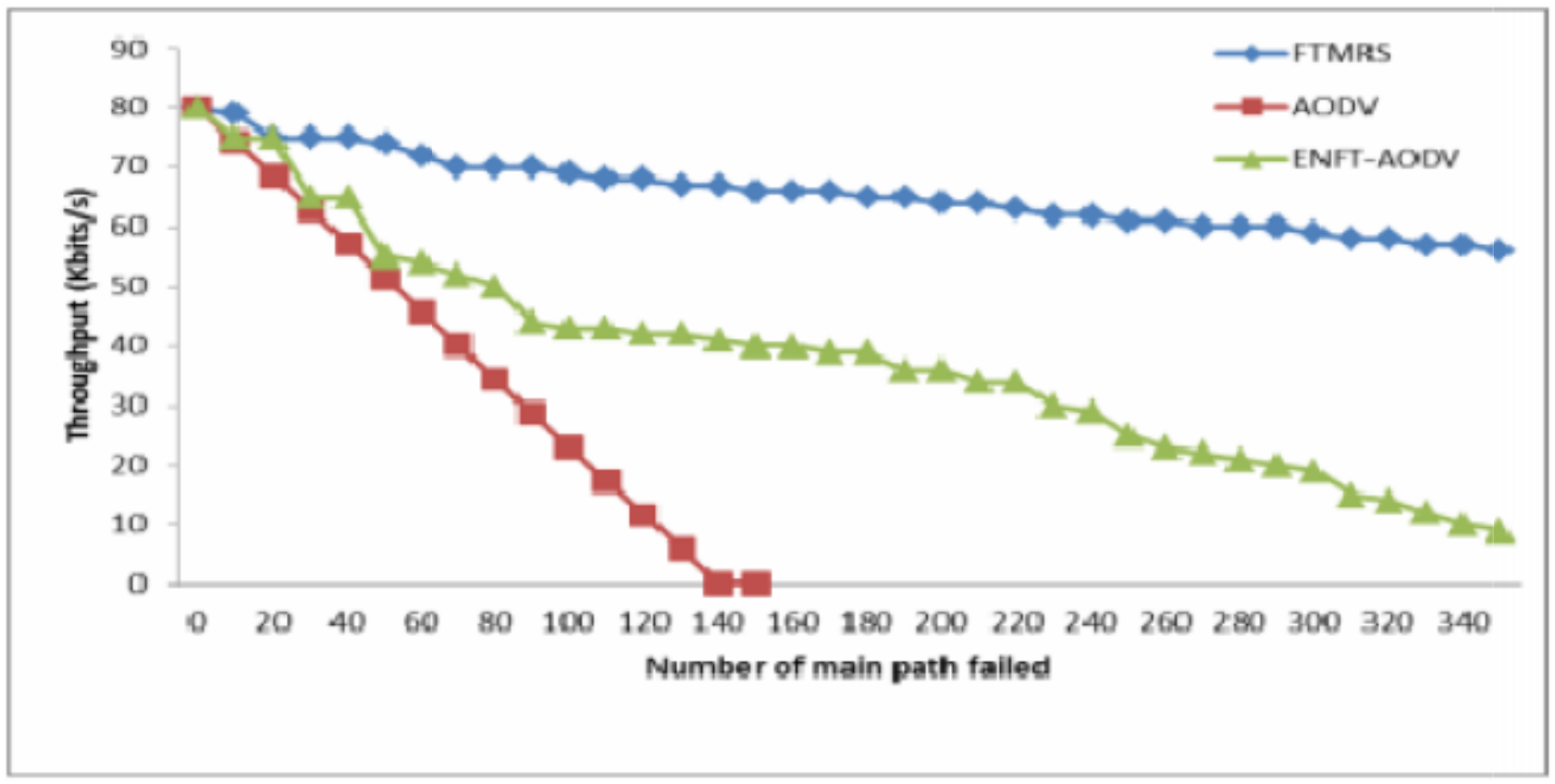

**Figure 9:** Throughput with respect to main routing path fault

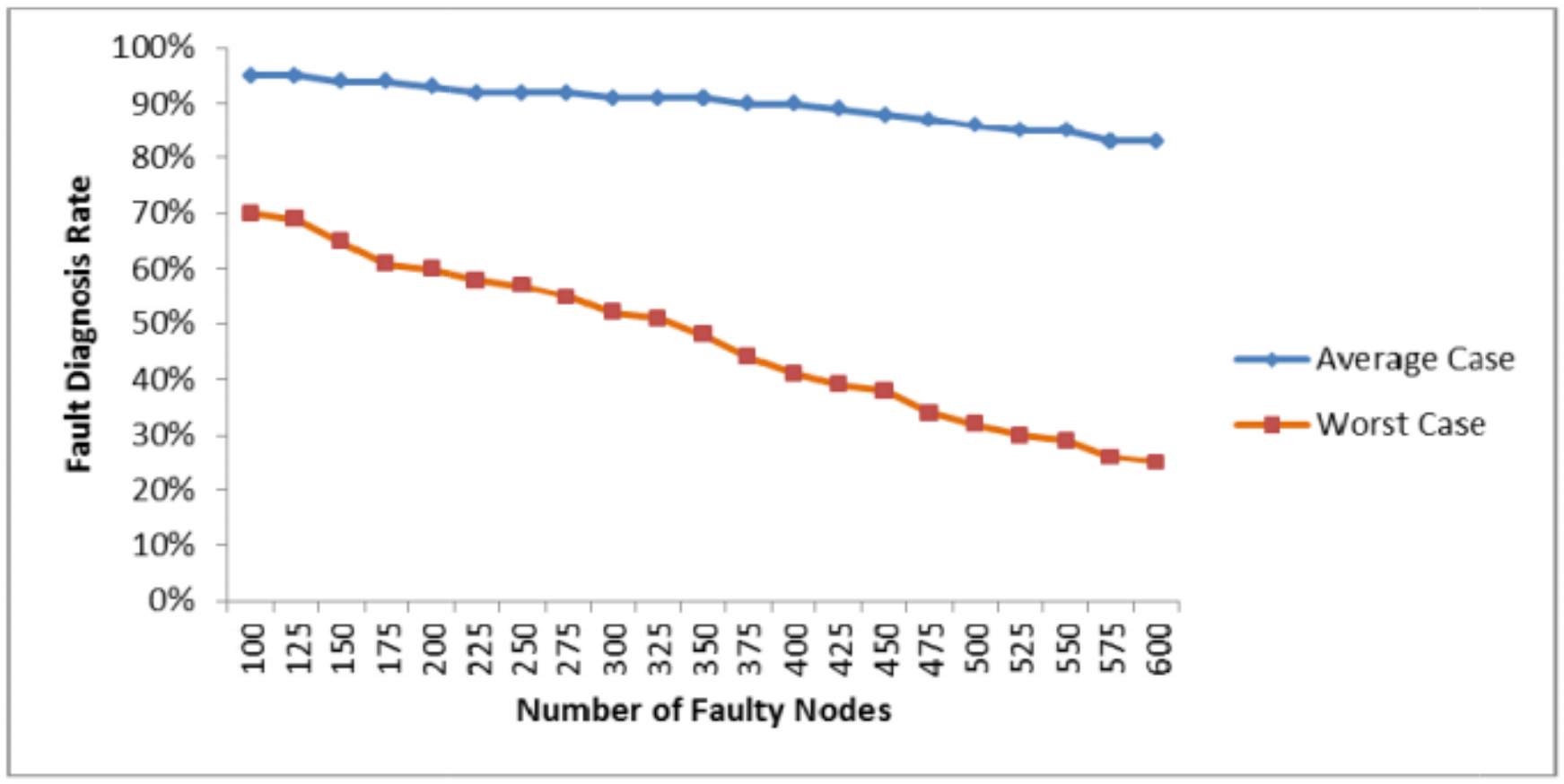

**Figure 10:** Successful diagnosis rate

Figure10 indicates the variation of the fault diagnosis rate with the number of faulty nodes. The fault diagnosis rate manifests the number of faulty node detected in each iteration. In average case if 100number of nodes are faulty,then approximately 95% of the faulty nodes are identified (95 out of 100), whereas approximately 70% of faulty nodes are detected (70 out of 100) in the worst case. In worst case high node failure in network leads to low fault diagnosis rate.

## 6. CONCLUSIONS

In this paper, we present FTMRS as a fault tolerant multipath routing scheme for energy efficient WSN. The FTMRS technique recovers node fault and transmission fault and transmits data in energy efficient manner. In FTMRS technique, fault tolerant percentage is very high compare to other fault tolerant techniques. Data routing time in FTMRS is very fast and energy aware even at high percentage of nodes fault. The FTMRSalso proposes a faulty node recovery scheme that effectively reuses or replace the faulty node. The simulation results establish that the proposed routing give better monitoring of the nodes that effectively leads to an energy efficient maximally fault tolerant in sensor network.

In future we would like to improve and analyze the time complexity of the proposed algorithm. Moreover, the performance in worst case scenario improved by efficient detection of faulty nodes.

## ACKNOWLEDGMENTS

This work is supported by the Council of Scientific and Industrial Research (CSIR) Human resource Development group (Extramural Research Division).

## Authors

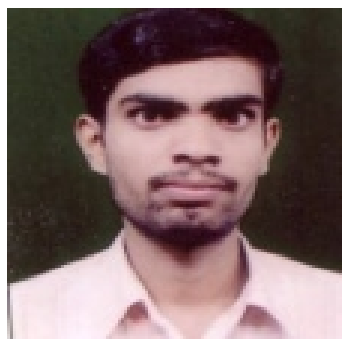

**PrasenjitChanak** received the M.Tech degree in information technology from Bengal Engineering and Science University (BESU), Shibpur, West Bengal, India in 2011 and is currently working toward the Ph.D. degree in Information Technology at Bengal Engineering and Science Univeisity, Shibpur, West Bengal. His main research interests are wireless ad hoc and sensor network.

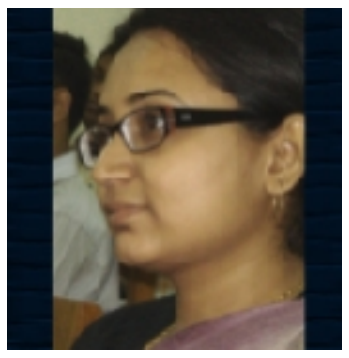

Dr. TuhinaSamanta is presently an assistant professor in Bengal Engineering and Science University, Shibpur. She completed her B.Tech and M.Tech from the Institute of Radiophysics and Electronics, Calcutta University in 2003 and 2005 respectively. She was awarded Canodia Research Scholarship during her M.Tech. She received her doctoral of philosophy degree from Bengal Engineering and Science University, Shibpur, in January, 2010.
Her major research areas are study and design of algorithms for VLSI Physical Design, Physical Design for Biochip and CAD, and development of algorithm in wireless sensor network. She has several publications in IEEE/ACM conferences and journals. She received Best paper award in "Computational method and software for contribution to algorithms, data structures, system architectures, software development for mechatronic and embedded systems" at 2010 IEEE/ASME International Conference on Mechatronic and Embedded Systems and Applications, July 15-17, 2010, China.

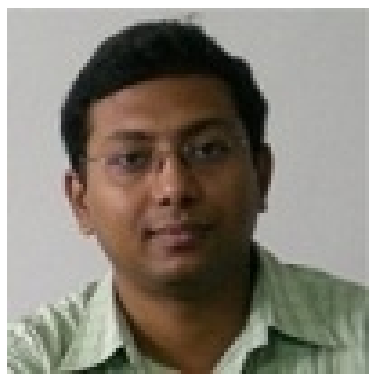

**Indrajit Banerjee** is an assistant professor in the Information Technology Department at Bengal Engineering and Science University, Shibpur, India. He got the bachelor degree in mechanical engineering from Institute of Engineers, India. He received his masters in Information Technology from Bengal Engineering and Science University in 2004. He is currently pursuing his Ph. D. in Information Technology in Bengal Engineering & Science University. His main research interests are cellular automata, wireless ad hoc and sensor network.